\documentclass[aps,pre,preprint,12pt,altaffilletter,showpacs]{revtex4}
\usepackage{graphicx}
\usepackage{amssymb}
\usepackage{amsmath} 
\usepackage{hyperref}
\usepackage{textcomp}
\usepackage{psfrag}
\usepackage{setspace}
\usepackage{url}
\begin{document}
\title{Effective medium theory of elastic waves in random networks of rods}
\author{J.~I.~Katz$^*$}
\affiliation{Department of Physics, Washington University, St.~Louis,
Mo.~63130}
\affiliation{McDonnell Center for the Space Sciences, Washington
University, St.~Louis, Mo.~63130}
\email{katz@wuphys.wustl.edu}
\author{J.~J.~Hoffman}
\author{M.~S.~Conradi}
\author{J.~G.~Miller}
\affiliation{Department of Physics, Washington University, St.~Louis,
Mo.~63130}
\date{}
\begin{abstract}
We formulate an effective medium (mean field) theory of a material
consisting of randomly distributed nodes connected by straight slender rods,
hinged at the nodes.  Defining novel wavelength-dependent effective elastic
moduli, we calculate both the static moduli and the dispersion relations of
ultrasonic longitudinal and transverse elastic waves.  At finite wave vector
$k$ the waves are dispersive, with phase and group velocities decreasing with
increasing wave vector.  These results are directly applicable to networks
with empty pore space.  They also describe the solid matrix in two-component
(Biot) theories of fluid-filled porous media.  We suggest the possibility of
low density materials with higher ratios of stiffness and strength to
density than those of foams, aerogels or trabecular bone.
\end{abstract}
\maketitle
\section{Introduction}
Many materials of biomedical or technological interest \cite{GA82} contain a
network of slender elastic rods with hinged connections at irregularly or
randomly distributed nodes.   In some cases the material may consist only of
the network, while in others, such as trabecular bone, the space between the
rods may be filled with fluid.  Calculation of the elastic and acoustic
properties of such a random network requires a statistical model.  There
have been extensive studies of fluid-filled networks 
(\cite{B56a,B56b,B62,B80,WL82,A83}) but these have generally concentrated
on the interaction between the matrix and the pore fluid, while treating
the matrix as a simple homogenous material.

The mechanical properties of networks of elastic rods are sensitive
functions of their geometry.  For example, simple cubic lattices have bulk
moduli that scale as the volumetric filling factor $\cal F$ of the rods but
have zero stiffness to shear along the lattice planes, while in FCC and BCC
lattices all moduli scale $\propto {\cal F}$.  The moduli of open-celled
foams scale $\propto {\cal F}^2$ while those of closed-cell foams scale
$\propto {\cal F}^3$ \cite{GA82}.  In general, if a deformation requires
extension or compression of rods the corresponding modulus scales $\propto
{\cal F}$, but if it is resisted only by rod flexure or by the resistance of
the nodes to changes in the rod angles the moduli scale as a higher power of
$\cal F$.  Foams are soft because they can deform by bending their thin
members without extension, changing angles at nodes or cell junctions with
little resistance.

Here we develop an elementary effective medium model of the elastic
properties of a statistically isotropic matrix consisting of thin elastic
rods joined by hinges at randomly distributed nodes.  This is surely an
oversimplified description of a real material, as any analytic model of a
random material must be, but may provide a useful guide to and
parametrization of its properties.  Statistical modeling is necessary
because the full three-dimensional structure of a solid consisting of
irregularly located and connected nodes is unlikely to be known
quantitatively and likely varies from realization to realization.  In a
biomaterial like cancellous bone it may vary from individual to individual,
and within a single bone in a single individual.  Even in statistically
homogeneous foams it depends on details of their preparation.

By connecting nodes of mean coordination number $C \ge D$, where $D$ is the
dimensionality, with straight rods that terminate on the nodes, we construct
a model in which (except for pathological cases like a simple cubic lattice)
strain implies changes in the distances between nodes that are first order
in the strain, and hence extensions of the rods of that order.  The
resulting structure is stiffer than open or closed cell foams such as
expanded polymers, aerogels, hydrogels, sponges and spongy (trabecular or
cancellous) bone in which rods or cell edges join at points along their
lengths \cite{GA82}.  In our model stiffness is derived from the resistance
of the rods to extension, rather than to bending; the former is much greater
for slender rods.  In contrast to structures near the isostaticity limit
\cite{BMLM10,BSM11,MSL11}, the nodes are overconstrained.  The great
stiffness of structures like those we model may have practical utility.

Within this model we calculate the dispersion of longitudinal and transverse
elastic waves at finite wave vectors, where heterogeneity is significant.
In this non-dissipative and non-local model dispersion is a consequence and
measure of the inhomogeneous spatial structure (nonlocality invalidates
one of the assumptions of the Kramers-Kronig relations).  The dispersion
relations, functions of a single scalar variable (the magnitude of the
wave-vector), are a compact description of a complex three-dimensional
structure. 

\section{The Model}
The model consists of nodes randomly, but statistically uniformly, 
distributed in space with a mean density $n$.  In contrast to the related
problems of the description of vibrations in a solid in which massive atoms
are connected by massless elastic rods or springs \cite{AM76}, or a granular
material in which grains are pressed into frictional contact \cite{ZSN08},
in our model the nodes are massless and serve only as geometric constraints
on the rods connecting them.  Each node is connected to an average of $C$
(the coordination number) nearest neighbor nodes by uniform and identical
(except for length) elastic rods that are hinged at the nodes.  We expect
the discrete random structure to approach the continuum (effective medium)
model as $C \to \infty$.  We present results for $C = 4$ and $C = 12$ to
show the dependence of the model on $C$.  Even $C = 4$ is well above the
matrix's isostaticity threshold \cite{BMLM10,BSM11,MSL11}, and is sufficient
to assure its stiffness against shear.

The distribution of distances $\ell_i$ to the $i$-th nearest neighbor of a
node is obtained from an Erlang distribution (Gamma distribution for integer
index) for the volume $V_i$ enclosed by a sphere of radius $\ell_i$
containing its $i$ nearest neighbors:
\begin{equation}
P_V(V_i) = {V_i^{i-1} n^i \exp{(-nV_i)} \over \Gamma(i)}.
\end{equation}
The probability distribution of $\ell_i$ is
\begin{equation}
P_\ell(\ell_i) = 4 \pi \ell_i^2 P_V(V_i),
\end{equation}
where $\ell_i = (3 V_i /4 \pi)^{1/3}$.

Fig.~1 shows one realization of a network of rods in a unit
cube connecting randomly distributed nodes.  The nodes have a mean density
$n = 250$ and mean coordination number $C = 12$.
\begin{figure}
\begin{center}
\includegraphics[width=5in]{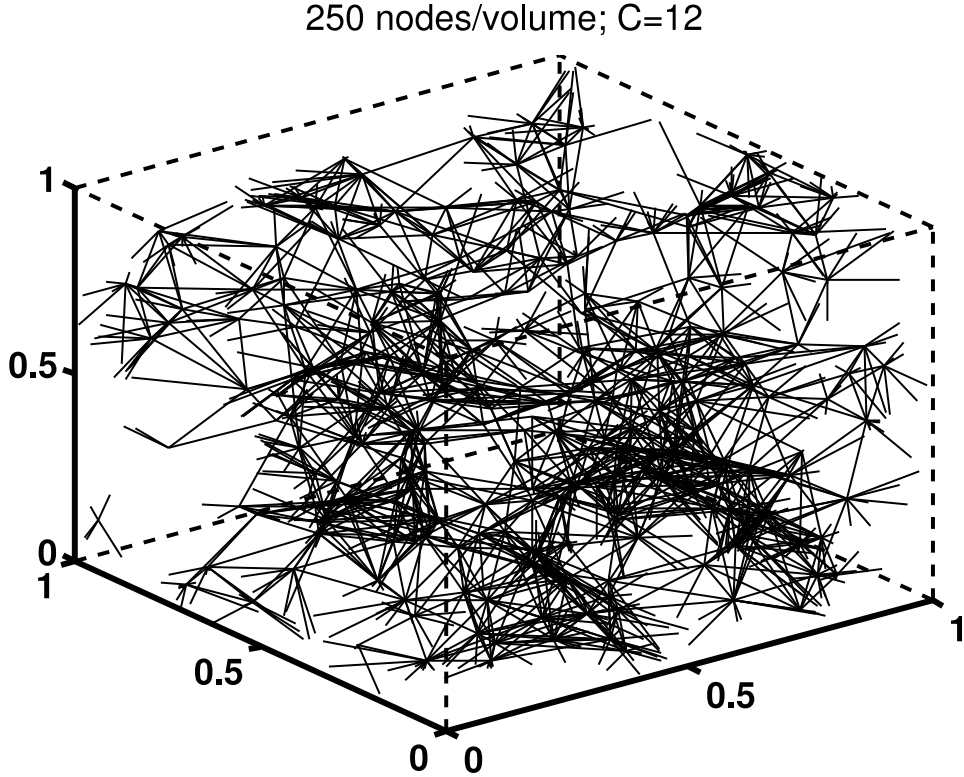}
\end{center}
\caption{Rods connecting randomly distributed nodes.  One realization of
randomly distributed nodes with mean density 250 within a unit cube is
shown.  The closest nodal neighbors are connected by rods, for a mean
coordination number $C = 12$; some of these are outside the cube.  The
network is characterized by a length $n^{-1/3} = 0.159$ and is not fractal.
The complexity of spatial structure, including voids and tongues of multiply
connected nodes, is evident, but difficult to quantify except by its
mechanical properties.}
\label{network}
\end{figure}

In a microscopic realization of randomly distributed nodes, such as that
shown in Fig.~1, it is not possible to enforce the condition that all nodes
have the same coordination number; if node $j^\prime$ is the $C$-th nearest
neighbor of node $j$, to which it is connected, node $j$ may (for example)
be the $C + 1$-st nearest neighbor of node $j^\prime$.  Such complications
are ignored, by definition, in an effective medium theory, in which $C$ may
be regarded as a {\it mean} coordination number.

We consider the dispersion relations of elastic waves in this medium of thin
straight massive elastic rods connecting massless nodes, with the rods 
completely hinged (free to change direction and to rotate about their axes)
at the nodes.  The effective medium model describes the motion of the
nodes by a sinusoidal plane wave.  The latter approximation is valid in the
limit $k n^{-1/3} \to 0$, but breaks down for short waves with $k n^{-1/3}
\gtrsim \pi$, for which scattering and localized modes are also important
\cite{C88,F96}.  In the limit of small amplitude (an assumption made
throughout) compressive loads are below the rods' Euler buckling thresholds.

The fundamental equations for a small amplitude plane longitudinal wave
propagating in the $x$ direction in an infinite medium are 
\begin{align}
{\partial v_x \over \partial t} &= - {1 \over \rho_0} {\partial \sigma_{xx}
\over \partial x} \\
{\partial u_{xx} \over \partial t} &= - {\partial v_x \over \partial x},
\end{align}
where $\sigma_{xx}$ is the $xx$ component of the stress tensor, $u_{xx} \ll
1$ is the $xx$ component of the strain tensor describing the elastic wave,
$v_x$ is the $x$ component of material velocity and $\rho_0$ is the mean
density (including void space) of the matrix.  In an effective medium model
we consider $\sigma_{xx}$, $u_{xx}$ and $v_x$ to be continuous functions of
space, as they would be in a continuous medium.
The wave variables $v_x/c_0$ and $\sigma_{xx}/{\cal B}_l(k)$ are of the same
order of smallness as $u_{xx}$, where $c_0$ is the thin fiber longitudinal
sound speed of the rods and ${\cal B}_l(k)$ is the effective longitudinal
modulus (defined below) of the matrix.  

These equations are closed with a constitutive relation
\begin{equation}
\label{constit}
\sigma_{xx} = {\cal B}_l(k) u_{xx},
\end{equation}
where the effective longitudinal modulus ${\cal B}_l(k)$ is independent of
space in an effective medium theory, but depends on the mechanical
properties of the rods and is a nontrivial function of the their spatial
structure and of the wave vector $k$ (a scalar in an isotropic effective
medium).  By taking ${\cal B}_l(k)$ to be real the model is lossless by
assumption.  A dissipative model could be defined by taking ${\cal B}_l(k)$
to be complex or, equivalently, by adding a term ${\cal B}^\prime_l(k) 
{\dot u}_{xx}$ to Eq.~\ref{constit}.

The modulus is defined by the relation
\begin{equation}
{\cal E}_{el} = {1 \over 2} {\cal B}_l(k) u_{xx}^2,
\end{equation}
where the elastic energy density ${\cal E}_{el}$ is obtained by calculating
the elastic energy added to the rods by the strain field $u_{xx}(x,t)$.

The resulting wave equation
\begin{equation}
{\partial^2 v_x \over \partial t^2} - {{\cal B}_l(k) \over \rho_0}
{\partial^2 v_x \over \partial x^2} = 0
\end{equation}
has solutions
\begin{equation}
v_x = v_{x0} \exp{i(k x \pm \omega t)},
\end{equation}
with the dispersion relation
\begin{equation}
\label{disper}
v_{ph} = {\omega \over k} = \sqrt{{\cal B}_l(k) \over \rho_0}.
\end{equation}

For infinitesimal displacements of a hinged straight rod's endpoints it
remains straight and below its buckling limit, so that its elastic energy is
that of stretching.  For slender rods, bending and torsional energy may be
ignored even if the nodes are not hinged.  Taking a cross-section $A$,
Young's modulus $E$ and equilibrium length $\ell$, stretched by an amount
(not positive-definite) $\Delta\ell$ satisfying $\vert\Delta\ell\vert \ll
\ell$, the elastic energy is
\begin{equation}
{\cal E}_{rod} = {(\Delta\ell)^2 A E \over 2 \ell}.
\end{equation}
We describe a longitudinal elastic wave of amplitude $a$ propagating in an
infinite medium in the $x$ direction with wave vector $k {\hat x}$ by a
displacement field
\begin{equation}
u_x(x) = a \sin{(k x + \zeta)},
\end{equation}
corresponding to a strain field
\begin{equation}
\label{uxx}
u_{xx} = a k \cos{(k x + \zeta)}.
\end{equation}
The difference in $x$-displacements of the ends of a rod, one end of which
defines $x = 0$ and the other is at $x = \ell \cos\theta + \delta u_x$, 
where $\theta$ is the angle between the unstrained rod and the $x$ axis, is
\begin{equation}
\label{ux}
\delta u_x = a[\sin{(k \ell \cos\theta + \zeta)} - \sin\zeta];
\end{equation}
$\delta u_y = \delta u_z = 0$.  The choice of one end of the rod as defining
$x = 0$ is equivalent to a choice of phase $\zeta$; either may be chosen
freely if, in the computation of total elastic energy and modulus, an
average is taken over the other.

The rod undergoes a change in length (to first order in $a$)
\begin{align}
\label{deltaell}
\Delta\ell &= \sqrt{\ell^2\sin\theta^2 + (\ell\cos\theta - \delta u_x)^2}
- \ell \\
&\approx \cos\theta \delta u_x \\
&\approx a \cos\theta [\sin{(k \ell \cos\theta + \zeta)} - \sin\zeta].
\end{align}
Its elastic energy is
\begin{equation}
\label{Elrod}
{\cal E}_{rod} = {A E \over 2 \ell} \cos^2\theta \delta u_x^2 = {A E \over
2 \ell} a^2 \cos^2\theta \left[\sin{(k \ell \cos\theta + \zeta)} - 
\sin\zeta\right]^2.
\end{equation} 
The mean elastic energy per volume is
\begin{equation}
\label{Elden}
{\cal E}_{el} = n {C \over 2} {A E a^2 \over 2} \left\langle{\cos^2\theta
\over \ell} \left[\sin{(k \ell \cos\theta + \zeta)} - \sin\zeta\right]^2
\right\rangle_{\zeta,\theta,\ell} \equiv {1 \over 2} {\cal B}_l(k) 
\left\langle u_{xx}^2 \right\rangle,
\end{equation}
where the factor of $n {C \over 2}$ allows for the presence of two rods per
node of coordination number $C$.

For the sinusoidal wave (\ref{uxx})
\begin{equation}
\langle u_{xx}^2 \rangle = {1 \over 2} a^2 k^2
\end{equation}
and
\begin{equation}
\label{bl}
{\cal B}_l(k) = n C {A E \over k^2} \left\langle {\cos^2\theta \over \ell}
\left[\sin{(k \ell \cos\theta + \zeta)} - \sin\zeta\right]^2
\right\rangle_{\zeta,\theta,\ell}.
\end{equation}
The dispersion relation (\ref{disper}) becomes
\begin{equation}
\label{displ}
\omega = \sqrt{{2 E \over \langle \ell \rangle \rho_m} \left\langle
{\cos^2\theta \over \ell} \left[\sin{(k \ell \cos\theta + \zeta)} - 
\sin\zeta\right]^2 \right\rangle_{\zeta,\theta,\ell}},
\end{equation}
where $\rho_0 = n {C \over 2} A \langle \ell \rangle \rho_m$ and $\rho_m$ is
the material density of the rods.  The coordination number enters only
through the average over $\ell$, which is weakly dependent on $C$ at finite
$k n^{-1/3}$.

The long wavelength limit is:
\begin{equation}
\label{blstat}
\lim_{k \to 0} {\cal B}_l(k) \to {{\cal F} \over 5} E,
\end{equation}
independent of $C$, where the volumetric filling factor 
\begin{equation}
{\cal F} \equiv {\rho_0 \over \rho_m} = {1 \over 2} n C A \langle \ell
\rangle.
\end{equation}
The dispersion relation in this limit is
\begin{equation}
\label{displstat}
{\omega \over k} \to \sqrt{E \over 5 \rho_m} = {v_{rod} \over \sqrt{5}},
\end{equation}
where $v_{rod} \equiv \sqrt{E/\rho_m}$ is the phase and group velocity of
the longitudinal wave in an individual rod in the (non-dispersive) thin-rod
limit.

The derivation has assumed ${\cal F} \ll 1$, a condition implicit in the
description of the matrix as a network of long slender straight rods with
$\ell \gg A^{1/2}$.  If this condition is not met, then the assumption that
the rods do not intersect between the pre-determined nodes is invalid.

\section{Transverse Dispersion Relation}
The dispersion relation of transverse waves is found from an analogous
calculation.  Defining $y$ as the polarization direction, we replace
$u_x$ by $u_y$, $v_x$ by $v_y$, $u_{xx}$ by $u_{xy}$ and $\sigma_{xx}$ by
$\sigma_{xy}$.  In Eqs.~\ref{deltaell}, \ref{Elrod}, \ref{Elden}, \ref{bl}
and \ref{displ}, outside the brackets $\sin\theta$ is replaced by
$\cos\theta$ and $\cos\theta$ by $\sin\theta\,\cos\phi$, where $\phi$ is the
azimuthal angle of the rod, taking $\phi = 0$ in the $x$-$y$ plane.
Averaging over $\phi$ introduces additional factors of \textonehalf.  The
results Eqs.~\ref{blstat}, \ref{displstat} for the static limit become
\begin{align}
\lim_{k \to 0} {\cal B}_t(k) &\to {1 \over 15} {\cal F} E \\
{\omega \over k} &\to \sqrt{E \over 15 \rho_m} = \sqrt{1 \over 15}
v_{rod}.
\end{align}

\section{Results}
The dispersion relations for the longitudinal and transverse modes are shown
in Fig.~2 for $C = 4$ and $C = 12$.  
\begin{figure}
\begin{center}
\includegraphics[width=4in]{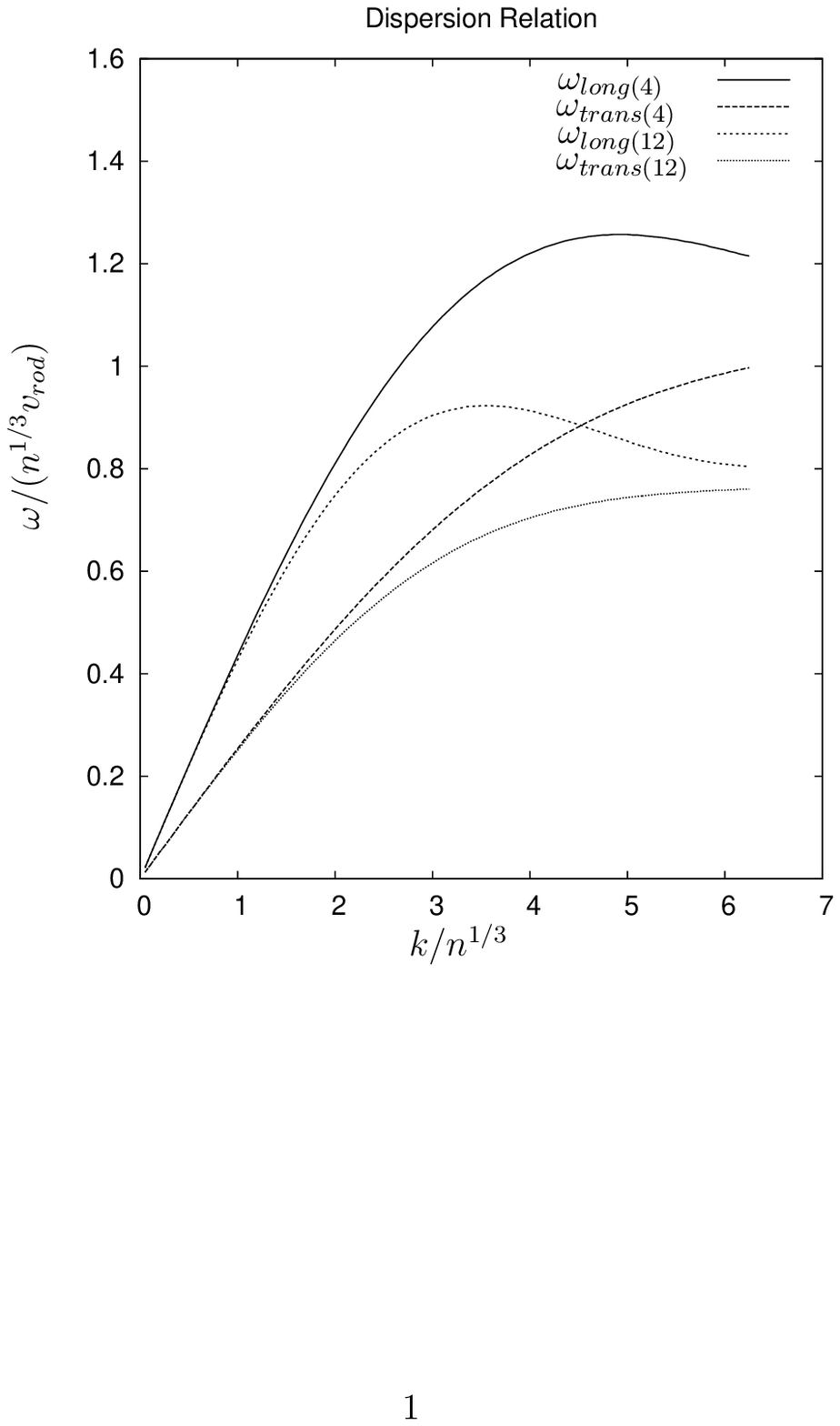}
\end{center}
\caption{Dispersion relations $\omega(k)$ of longitudinal and transverse
waves.  $v_{rod} \equiv \sqrt{E/\rho_m}$ is the slender rod
longitudinal wave propagation speed of the individual rods making up the
matrix.  The effective medium theory is expected to break down for $k 
n^{-1/3} \gtrsim \pi$ where scattering is strong.  At finite $k n^{-1/3}$
these dispersion relations depend on the coordination number, as shown for
$C = 4,\,12$.  Dispersion is greater for larger $C$ because of the presence
of longer rods with larger values of $k \ell$; for $C = 4$, $\langle \ell
\rangle n^{1/3} = 0.778$ while for $C = 12$, $\langle \ell \rangle n^{1/3} =
1.085$.}
\label{disperfig}
\end{figure}
The frequency dependences of the corresponding phase and group wave
velocities $v_{ph}$ and $v_{gr}$ are shown in Fig.~3.
\begin{figure}
\begin{center}
\includegraphics[width=4in]{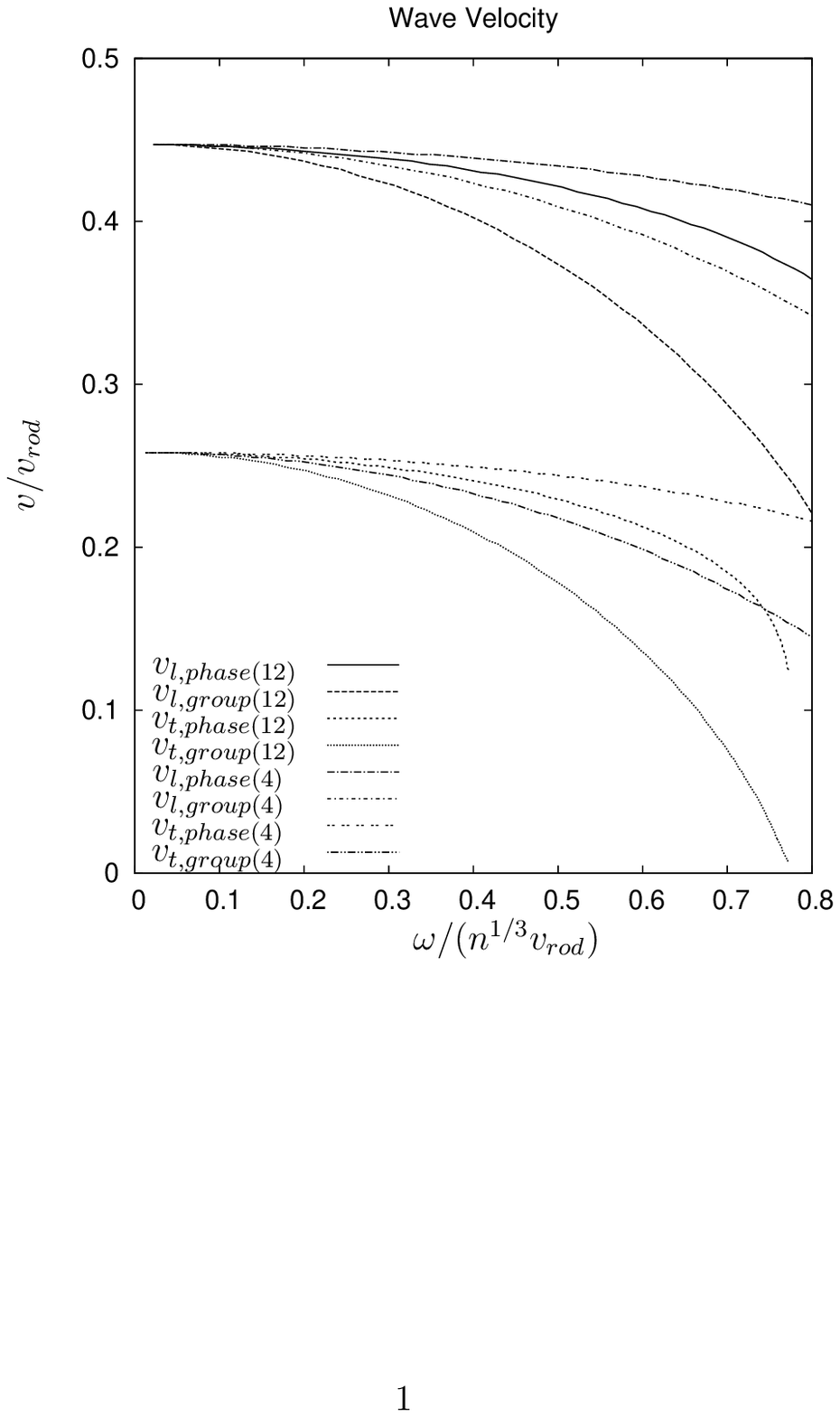}
\end{center}
\caption{Phase and group velocities of longitudinal and transverse waves as
functions of frequency. $C$ is indicated in subscripts.  The effective medium
theory is expected to break down for $\omega /(n^{1/3} v_{rod}) \gtrsim 1$.
At finite $k n^{-1/3}$ these velocities depend on the coordination number.}
\label{velfig}
\end{figure}

From the preceding results for the longitudinal and transverse wave speeds
the static ($\omega,k \to 0$) Young's modulus $E_{matrix}$ and Poisson's
ratio $\nu_{matrix}$ of the bulk matrix may be obtained \cite{LL59}:
\begin{align}
\label{moduli}
E_{matrix} &={1 \over 6} {\cal F} E \\
\nu_{matrix} &= {1 \over 4}.
\end{align}
The model predicts the ratio $v_{tr}/v_{long} = 1/\sqrt{3}$ of long
wavelength ($\lambda \gg n^{-1/3}$) wave speeds, independent of the
properties of the rods.

For comparison, we present results for the elastic constants of cubic
lattices of rods and for the propagation speeds along the crystal axes of
long-wavelength waves in Table~\ref{cubic}.
\begin{table}
\begin{center}
\begin{tabular}{|c|c|c|c|}
\hline
\qquad&\ Face Centered Cubic \ &\ Body Centered Cubic \ & \ Simple Cubic \ \\
\hline
 $C_{11}$ & ${1 \over 6} {\cal F} E$ & ${1 \over 9} {\cal F} E$ &
 ${1 \over 3} {\cal F} E$ \\
 $C_{12}$ & ${1 \over 12} {\cal F} E$ & ${1 \over 9} {\cal F} E$ & 0 \\
 $C_{44}$ & ${1 \over 12} {\cal F} E$ & ${1 \over 9} {\cal F} E$ & 0 \\
 $v_{long}$ & ${1 \over \sqrt{6}} v_{rod}$ & ${1 \over 3} v_{rod}$ &
 ${1 \over \sqrt{3}} v_{rod}$ \\
 $v_{tr}$ & ${1 \over \sqrt{12}} v_{rod}$ & ${1 \over 3} v_{rod}$ & 0 \\
\hline
\end{tabular}
\end{center}
\caption{Elastic constants (from Eq.~\ref{Elden}) and wave speeds for
propagation along crystal axes for cubic lattices of rods.}
\label{cubic}
\end{table}

The acoustic properties of fluid-filled porous media, such as aquifers,
petroleum reservoirs and cancellous (trabecular) bone, are described by
models originally developed by Biot \cite{B56a,B56b,B62,B80,WL82,A83}.
These are also effective medium models that describe the wave modes of the
fluid filled medium as if it were homogeneous and continuous, and do not
attempt to include the dispersion at finite $k$, other than that which is a
consequence (by the Kramers-Kronig relations \cite{OJM78,J99,WMM05}) of
dissipation.  The properties of the empty matrix, such as those we have
modeled, are required to calculate the properties of the Biot modes.

We have found elastic moduli (Eq.~\ref{moduli} and Table~\ref{cubic}) that
are proportional to the density or filling factor.  Our model shows how
stiff low density structures can be, if designed or evolved to maximize
stiffness.  Our model describes a random structure; similar stiffness has 
been found for ordered structures \cite{S11}.  This is in contrast to
aerogels \cite{MRPJS00,MMREM01,LLL11}
that typically show $E \propto \rho_0^m$ with $2 \lesssim m \lesssim 4$.
These values are characteristic of closed-cell foams for which $m = 3$
\cite{GA82}, although one study of aerogels \cite{RBHP06} found $m
\approx 1.5$.  The structures of aerogels are complex and sensitive to their
preparation \cite{MMREM01,RBHP06,RMJ11,OWL11,LLL11}.  

A study of water-filled trabecular bone \cite{TE91} found results that can
be fitted by a value for the matrix of $m \approx 3.7$.  These properties
are very different from those of our model, and mechanical stiffness cannot
have driven the evolution of trabecular bone.  The remarkably low modulus
of the trabecular matrix explains the observation \cite{PPL09} that at low
matrix filling factor the fluid properties largely determine the elastic
wave speed and attenuation in trabecular bone.  This may explain why Biot
models of fluid-filled trabecular bone have required phenomenological
fitting of their parameters \cite{W92,HO97,H99,HL99,MSM03}.

In our effective medium model dispersion is a consequence of microstructure
and is present even without dissipation.  The apparent contradiction with
the Kramers-Kronig relations as adopted in \cite{OJM78,J99,WMM05} is
resolved, as it is in textbook point mass and spring models of phonons in
crystals \cite{AM76}, by noting that the effective medium model, like the
phonon models, is nonlocal: the force on a mass is determined by the
instantaneous positions of other masses.  This violates the Kramers-Kronig
assumption of locality, that response at one point is determined only by
the history of fields at that point\footnote{It might be imagined that if
the response were instantaneous but nonlocal, relations analogous to the
Kramers-Kronig relations could be developed as functions of the spatial
wavenumber.  However, the response at any point would depend on the stimulus
in all directions, so there is no spatial analogue of causality.}.  The
mechanical compliance, analogous to the dielectric permittivity, is here a
function of both $\omega$ and $k$.  Our model makes first-principles
quantitative predictions of the dispersion and elastic wave speeds of
vacuum-filled matrices that may be tested experimentally.

\begin{acknowledgments}
We thank P.~V.~Bayly, L.~M.~Canel-Katz, A.~E.~Carlsson, G.~M.~Genin,
M.~Holland, B.~Johnson, M.~Milne, N.~D.~Mermin, A.~Nelson, D.~R.~Nelson and
Z.~Nussinov for unpublished results and discussions.  This work was
supported in part by NIH grant R01AR057433, by grant P30AR057235 from the
National Institute of Arthritis and Musculoskeletal and Skin Diseases to the
Washington University Core Center for Musculoskeletal Biology and Medicine
and by the American Chemical Society Petroleum Research Fund \#51987-ND9.
\end{acknowledgments}


\begin{thebibliography}{99}
\bibitem{GA82}L.~J.~Gibson, and M.~F.~Ashby,
{\it Proc.~R.~Soc.~Lond.\/} A {\bf 382}, 43--59 (1982).
\bibitem{B56a}M.~A.~Biot, 
{\it J.~Acoust.~Soc.~Am.\/} {\bf 28}, 168--178 (1956).
\bibitem{B56b}M.~A.~Biot, 
{\it J.~Acoust.~Soc.~Am.\/} {\bf 28}, 179--191 (1956).
\bibitem{B62}M.~A.~Biot,
{\it J.~Acoust.~Soc.~Am.\/} {\bf 34}, 1254--1264 (1962).
\bibitem{B80}J.~G.~Berryman,
{\it Appl.~Phys.~Lett.\/} {\bf 37}, 382--384 (1980).
\bibitem{WL82}J.~L.~Williams, and J.~L.~Lewis, 
{\it J.~Biomech.~Eng.\/} {\bf 104}, 50--56 (1982).
\bibitem{A83}K.~Attenborough,
{\it J.~Acoust.~Soc.~Am.\/} {\bf 73}, 785--799 (1983).
\bibitem{BMLM10}C.~P.~Broedersz, X.~Mao, T.~C.~Lubensky, and 
F.~C.~MacKintosh, 
{\it Nature Physics\/} {\bf 7}, 983--988 (2011).
\bibitem{BSM11}C.~P.~Broedersz, M.~Sheinman, and F.~C.~MacKintosh, 
{\it Phys.~Rev.~Lett.\/} {\bf 108}, 078102 (2012)
(\url{http://arXiv.org/abs/1108.4354}).
\bibitem{MSL11}X.~Mao, O.~Stenull, and T.~C.~Lubensky, 
{\it Physical Review E\/} in press (2012)
(\url{http://arXiv.org/abs/1111.1751}).
\bibitem{AM76}N.~W.~Ashcroft and N.~D.~Mermin {\it Solid State Physics\/}
(Holt, Reinhart and Winston, New York, 1976) pp.~422--447.
\bibitem{ZSN08}Z.~Zeravcic, W.~van~Saarloos, and D.~R.~Nelson
{\it Europhys.~Lett.\/} {\bf 83}, 44001 (2008).
\bibitem{C88}C.~A.~Condat,
{\it J.~Acoust.~Soc.~Am.\/} {\bf 83} (2), 441--452 (1988).
\bibitem{F96}M.~Foret, E.~Courtens, R.~Vacher, and J.~B.~Suck,
{\it Phys.~Rev.~Lett.\/} {\bf 77}, 3831--3834 (1996).
\bibitem{LL59}Landau, L.~D., Lifschitz, E.~M. {\it Theory of
Elasticity\/} 1st ed. (Addison-Wesley, Reading, Mass., 1959), p.~99 Eq.~22.4.
\bibitem{OJM78}M.~O'Donnell, E.~T.~Jaynes, and J.~G.~Miller,
{\it J.~Acoust.~Soc.~Am.\/} {\bf 63}, 1935--1937 (1978).
\bibitem{J99}J.~D.~Jackson, {\it Classical Electrodynamics\/} 3rd ed. 
(Wiley, New York, 1999) pp.~330--335. 
\bibitem{WMM05}K.~R.~Waters, J.~Mobley, and J.~G.~Miller,
{\it IEEE Trans.~Ultrason.~Ferroelect.~Freq.~Control\/} {\bf 52}, 822--833 
(PMID:16048183) (2005).
\bibitem{S11}T.~A.~Schaedler, {\it et al.\/},
{\it Science\/} {\bf 334}, 962--965 (2011).
\bibitem{MRPJS00}H.~S.~Ma, A.~P.~Roberts, J.~H.~Pr\'evost, R.~Jullien,and
G.~W.~Scherer,
{\it J.~Non-Cryst.~Solids\/} {\bf 277}, 127--141 (2000).
\bibitem{MMREM01}M.~Moner-Girona, E.~Mart\'inez, A.~Roig, J.~Esteve, and
E.~Molins,
{\it J.~Non-Cryst.~Solids\/} {\bf 285}, 244--250 (2001)
\bibitem{RBHP06}A.~V.~Rao, S.~D.~Bhagat, H.~Hirashima, and G.~M.~Pajonk,
{\it J.~Colloid Interface Science\/} {\bf 300}, 279--285 (2006).
\bibitem{RMJ11}J.~P.~Randall, M.~A.~B.~Meador, and S.~C.~Jana, S.~C.
{\it ACS Applied Materials \& Interfaces\/} {\bf 3}, 613--626 (2011).
\bibitem{OWL11}K.~A.~D.~Obrey, K.~V.~Wilson, D.~A.~Loy, 
{\it J.~Non-Cryst.~Solids\/} {\bf 357}, 3435--3441 (2011).
\bibitem{LLL11}H.~Lu, H.~Luo, and N.~Leventis, in
{\it Aerogel Handbook\/} edited by M.~A.~Aegerter (Springer, New York,
2011) pp. 499--535.
\bibitem{TE91}M.~B.~Tavakoli, and J.~A.~Evans,
{\it Phys.~Med.~Biol.\/} {\bf 36}, 1529--1537 (1991).
\bibitem{PPL09}M.~Pakula, F.~Padilla, and P.~Laugier,
{\it J.~Acoust.~Soc.~Am.\/} {\bf 126}, 3301--3310 (2009).
\bibitem{W92}J.~L.~Williams, 
{\it J.~Acoust.~Soc.~Am.\/} {\bf 91}, 1106--1112 (1992).
\bibitem{HO97}A.~Hosokawa, and T.~Otani,
{\it J.~Acoust.~Soc.~Am.\/} {\bf 101}, 558--562 (1997).
\bibitem{H99}E.~R.~Hughes, {\it et al.\/},
{\it Ultrasound Med.~Biol.\/} {\bf 25}, 811--821 (1999).
\bibitem{HL99}T.~J.~Haire, and C.~M.~Langton, 
{\it Bone\/} {\bf 24}, 291--295 (1999). 
\bibitem{MSM03}M.~M.~Mohamed, L.~T.~Shaat, and A.~N.~Mahmoud, 
{\it IEEE Trans. Ultrasonics, Ferroelectrics, and Frequency Control\/}
{\bf 50}, 279--288 (2003). 
\end{thebibliography}
\end{document}